# Interlaced linear-nonlinear optical waveguide arrays

**Kyriakos Hizanidis, [1,*] Yannis Kominis, [1] and Nikolaos K. Efremidis[2]**

[1] *School of Electrical and Computer Engineering, National Technical University of Athens, Athens, Greece*
[2] *Department of Applied Mathematics, University of Crete, Crete, Greece*
*Corresponding author: kyriakos@central.ntua.gr*

**Abstract:** The system of coupled discrete equations describing a two-component superlattice with interlaced linear and nonlinear constituents is revisited as a basis for investigating binary waveguide arrays, such as ribbed AlGaAs structures, among others. Compared to the single nonlinear lattice, the interlaced system exhibits an extra band-gap controlled by the, suitably chosen by design, relative detuning. In more general physics settings, this system represents a discretization scheme for the single-equation-based continuous models in media with transversely modulated linear and nonlinear properties. Continuous wave solutions and the associated modulational instability are fully analytically investigated and numerically tested for focusing and defocusing nonlinearity. The propagation dynamics and the stability of periodic modes are also analytically investigated for the case of zero Bloch momentum. In the band-gaps a variety of stable discrete solitary modes, dipole or otherwise, in-phase or of staggered type are found and discussed.

## 1. Introduction

Periodic photonic structures in nonlinear dielectric media are a subject of intense theoretical and experimental research in our days. The formation of self-trapped localized solitary modes, among others, is of major importance.[1,2,3] Such solitons have eigenvalues inside the gaps of the band structure and their existence is due to the interplay between the effective lattice diffraction and nonlinearity. A major attribute of theirs, that facilitates their experimental observation[3], is their robustness under propagation. This attribute makes the periodic photonic structures ideal for applications in integrated photonic devices and waveguide arrays, such as multiport beam coupling, steering, and switching.[5–10] Furthermore, the related field of Bose–Einstein condensates loaded in optical lattices[11–15] is of increasing scientific interest and, thus, both fields progress in parallel as far as the theoretical investigation is concerned. The latter and, more specifically, the study of the formation and propagation of spatially localized modes in photonic structures, has been mostly based of either the tight-binding approximation or the coupled-mode theory, thus, rendering simplified discrete models.[16–20] However, these approximations provide accurate modeling only under the corresponding assumptions. A more general model is the nonlinear Schrödinger equation (NLSE) with spatially periodic coefficients, where the longest transverse dimension, $y$, has been integrated out by appropriate averaging, namely,

$$i\frac{\partial E}{\partial z} + \frac{1}{2\beta_0}\frac{\partial^2 E}{\partial x^2} + k_0 \Delta n(x) E + k_0 n_2(x)|E|^2 E = 0, \qquad (1)$$

where $E$ is the $y$-averaged envelope electric field intensity, $k_0 = 2\pi/\lambda$ ($\lambda$ is the wavelength of the carrier), $\beta_0 = k_0 <n(x)>$ ($<n(x)>$ is the mean value of the linear refractive index), $n_2(x)$ models the periodic nonlinear refractive index in $(m/V)^2$ and $\Delta n(x)$ models the periodic modulation of the linear refractive index. Such photonic structures where both linear and nonlinear refractive indices are transversely modulated are of increasing interest lately due to special properties they possess.[21] The particular cases of periodic functions in the form of periodic sequences of Dirac functions[2, 22-24] or piecewise-constant coefficients[25-27] (nonlinear Kronig-Penney model) have previously been considered. Apart of particular cases as the latter, Eq. (1) is not amenable to a straightforward and conclusive analytical treatment.

In the present work the discrete re-casting of Eq. (1) is revisited. The investigation is aiming towards a model of a binary optical waveguide array[28] amenable to a thorough analytical treatment and, thus, predictable and controllable as far as the functions of the conceptual device it models are concerned. The main incentive is the experimental observation of discrete gap solitons in such AlGaAs binary arrays[28]. Periodic modulation of both refractive indices can be conceptually achieved in ribbed AlGaAs waveguides. In Fig. 1 such a configuration is illustrated: The mole fraction of the aluminum in the substrate (*S*) is higher than the respective one in the guiding ribbed region (*W*); that leads to a relatively higher refractive index in the latter. The modulated rib structure, on the other hand, renders the effective linear refractive index in the ribs piece-wise constant with higher value ($n_L$) in the transversely (i.e., in *y*) and longitudinally (i.e., in *x*) wider segment and lower ($n_N$) in the transversely narrower one. However, there exist a geometric refractive index contrast between each rib and the region that separates them from the neighboring uneven ribs. Thus, light can be guided in the region under both the families of ribs. On the other hand, the effective nonlinear refractive index is also piece-wise constant attaining a considerably higher value in the narrower segment. Eventually, focusing will enable light to be efficiently guided in the small ribs while, at the same time, the much higher linear refractive index contrast of the large ribs will ensure the capability of guiding light by them too. However, relatively speaking, the wider segment can be considered almost linear as we will show shortly. Similar binary arrays can, conceptually at least, be thought in silica (of considerably lower nonlinearity than AlGaAs): A planar arrangement of two interlaced silica waveguides of relatively large and small diameters embedded in a silica cladding of slightly lower refractive index.

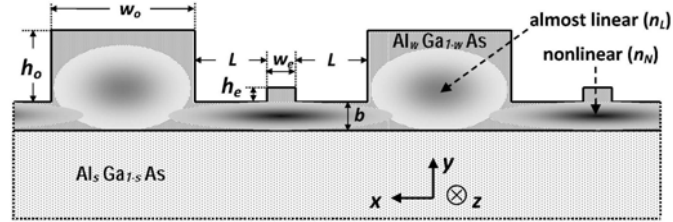

Fig. 1. A binary ribbed AlGaAs structure (*W* and *S* are Aluminum mole fraction respectively in the ribbed and the substrate regions). The shaded areas sketch coupled beams of light.

For a system such as the one illustrated in Fig. 1 and for optical propagation inside the first band of the band structure, Eq. (1) can readily be replaced by a pair of discrete nonlinear Schrödinger equations (DNLSEs) via a tight-binding approximation performed on the binary structure involved. By choosing even (odd) node-indices for the set of the relatively small (large) segments one readily obtains,

$$i\frac{dA_{2m}}{dz}+C\left(A_{2m-1}+A_{2m+1}\right)+V_{even}A_{2m}+G_{even}\left|A_{2m}\right|^2 A_{2m}=0$$
$$i\frac{dA_{2m\pm 1}}{dz}+C\left(A_{2m}+A_{2m\pm 2}\right)+V_{odd}A_{2m\pm 1}+G_{odd}\left|A_{2m\pm 1}\right|^2 A_{2m\pm 1}=0$$
(2)

where *C* is the coupling coefficient (in $m^{-1}$), the magnitude of $A_n$ (related to the power of the beam) is in $W^{1/2}$, $G_{even}$ and $G_{odd}$ are the respective nonlinear coefficients in $W^{-1}m^{-1}$ and $V_{even}$ and $V_{odd}$ are the respective de-tuning. For example, for TE (TM) polarization, $S=0.195$, $W=0.18$, $w_o=10\mu m$, $w_e=2\mu m$, $h_o=5\mu m$, $h_e=1\mu m$, $b=2\mu m$, $L=5\mu m$, $\lambda=1.55\mu m$ and with the nonlinear refractive index for AlGaAs considered as polarization independent and equal to $1.5\times10^{-13}$ $cm^2/W$, one obtains, $C=1.23$ $mm^{-1}$($1.22$ $mm^{-1}$), $V_{even}=10.10$ $mm^{-1}$ ($10.33$ $mm^{-1}$), $V_{odd}=12.75$ $mm^{-1}$ ($13.00$ $mm^{-1}$), $G_{even}=97.49$ $W^{-1}m^{-1}$($97.92$ $W^{-1}m^{-1}$), $G_{odd}=1.81$ $W^{-1}m^{-1}$ ($1.81$ $W^{-1}m^{-1}$); that is, $G_{odd}<<G_{even}$ and the DNLSE system, Eq. (2), for all practical purposes, is almost polarization independent. By normalizing the propagation distance via $\zeta=2Cz$ and applying the transformation, $A_m = \psi_m (2C/G_{even})^{1/2}\exp(i\zeta)$, one readily obtains,

$$i\frac{d\psi_{2m}}{d\zeta}+\frac{1}{2}(\psi_{2m-1}+\psi_{2m+1}-2\psi_{2m})+\frac{V_{even}}{2C}\psi_{2m}+|\psi_{2m}|^2\psi_{2m}=0$$
$$i\frac{d\psi_{2m\pm1}}{d\zeta}+\frac{1}{2}(\psi_{2m}+\psi_{2m\pm2}-2\psi_{2m\pm1})+\frac{V_{odd}}{2C}\psi_{2m\pm1}+\frac{G_{odd}}{G_{even}}|\psi_{2m\pm1}|^2\psi_{2m\pm1}=0$$
(3)

where, the nonlinear term in the second DNLSE can be ignored since $G_{odd}/G_{even}\approx1.8\%$ for the aforementioned example. The latter ratio is usually less than *1%* to *2%*, depending upon the geometrical details of the optical arrays in hand such as the ones based on ribbed AlGaAs configurations. Equations of the type of Eq. (3) are the subject matter of this work. This system of coupled discrete equations corresponds to a two-component superlattice that has also been previously proposed.[24]: It provides a generalization of the DNLS theory with a wider applicability than the tight-binding approximation and its importance has been explicitly stressed. It represents, therefore, a suitable discretization scheme for the continuous NLS model in media with transversely modulated linear and nonlinear. However, its apparent complexity (compared with previous single DNLS models) was considered in the past as a drawback for further pursuing it.

In the present work it is shown that, due to the effectively linear character of the second DNLSE (practically, a discrete linear Schrödinger equation), the system is actually amenable to analytical investigation of its behavior to a considerable extent. Thus, useful conclusions could be obtained as far as properties, functionality and controllability of devices of the type discussed are concerned. It must be emphasized that the investigation presented is not exhaustive since the problem in hand has a multitude of aspects which worth further investigation. However, the findings and the conclusions of the present study provide a firm analytical basis for this endeavor.

The paper is organized as follows: The model and the corresponding nonlinear diffraction relation along with the comparison of the transmission spectrum of propagating continuous wave modes with the single DNLS are presented in Section II. In Section III, the equation-condition for the modulational instability (MI) of these modes is derived and the latter are investigated at the base, the edge and inside the Brillouin zone as far as their stability and their characteristic features are concerned. In Section IV, the propagation dynamics of the continuous wave solutions of zero transverse momentum are analytically investigated and discussed. In Section V, numerically found solitary solutions are presented. Finally, in Section VI the main conclusions are summarized.

## 2. The model

In the following we consider Eq.(3) within a more general perspective, namely we allow for defocusing nonlinearity as well:

$$i\frac{d\psi_{2m}}{d\zeta}+\frac{1}{2}(\psi_{2m-1}+\psi_{2m+1}-2\psi_{2m})+\varepsilon_{even}\psi_{2m}+\sigma\psi_{2m}|\psi_{2m}|^2=0$$
$$i\frac{d\psi_{2m\pm1}}{d\zeta}+\frac{1}{2}(\psi_{2m\pm2}+\psi_{2m}-2\psi_{2m\pm1})+\varepsilon_{odd}\psi_{2m\pm1}=0$$
(4)

where $\sigma=\pm 1$. The phase shift parameters $\varepsilon_{even}$ and $\varepsilon_{odd}$, from the point of view of optical configurations as discussed in the Introduction are respectively the normalized detuning parameters $V_{even}/2C$ and $V_{odd}/2C$. In a broader perspective, however, they may vary in a wide range of values.

The nonlinear spatial dispersion (diffraction) condition for the existence of propagating modes can be found by assigning to $\psi_n$ the form $\psi_n=v_n\exp(ik_z\zeta)$

$$-(k_z+1-\varepsilon_{even})v_{2m}+\frac{1}{2}(v_{2m-1}+v_{2m+1})+\sigma v_{2m}|v_{2m}|^2=0$$

$$-(k_z+1-\varepsilon_{odd})v_{2m\pm 1}+\frac{1}{2}(v_{2m\pm 2}+v_{2m})=0$$

(5)This system renders the following expression for the even sites:

$$-\left[k_z+1-\varepsilon_{even}-\frac{1}{2(k_z+1-\varepsilon_{odd})}\right]v_{2m}+\frac{v_{2m+2}+v_{2m-2}}{4(k_z+1-\varepsilon_{odd})}+\sigma v_{2m}|v_{2m}|^2=0 \qquad (6)$$

for continuous (plain) wave solutions (CW) for the even-site lattice of the form $v_{2m}=A\exp(-imq)$, where the transverse (Bloch) momentum $q$ is $q=2k_xD$ with $D$ being the common spatial period of the lattice. Then, the odd-site lattice also supports CW solution of the form $v_{2m\pm 1}= v_{2m}(1+1/e^{\pm iq})/[2(\kappa+\Delta\varepsilon)]$, where $\Delta\varepsilon=\varepsilon_{even}-\varepsilon_{odd}$ is the normalized relative detuning (for optical configurations as such as in Fig. 1, usually negative) and $\kappa = k_z+1-\varepsilon_{even}$ is related to the propagation constant. In order for such periodic solutions to exist $|A|^2$ must be positive. Therefore, the following condition must hold:

$$\frac{\sigma}{2}\left(2\kappa-\frac{1+\cos q}{\kappa+\Delta\varepsilon}\right)=|A|^2\geq 0 \qquad (7)$$

Note that for a single nonlinear lattice of spatial period $2D$ the respective condition is fundamentally different,

$$\sigma(\kappa-\cos q)=|A|^2\geq 0 \qquad (8)$$

In Fig. 2 various examples for both cases and for both signs of $\sigma$ and $\Delta\varepsilon$ are presented. At the base ($q=0$) and the edge ($q=\pi$) of the Brillouin zone the group velocity, $d\kappa/dq$, vanishes. The respective $\Delta\varepsilon$-dependent pairs of $\kappa$'s (characterizing diatomic lattices) are $\Delta\varepsilon/2 \pm (\Delta\varepsilon^2/4+1)^{1/2}$ and $\Delta\varepsilon/2 \pm |\Delta\varepsilon/2|$. In contrast, the single nonlinear lattice of spatial period $2D$ the respective single $\kappa$'s are $1$ and $-1$. More importantly, there is a fundamental difference between the diatomic lattice and the single nonlinear lattice: There exist two transmission bands, namely: $max(-\Delta\varepsilon,0)0\leq\kappa\leq(\Delta\varepsilon^2/4+1)^{1/2}-\Delta\varepsilon/2$ and $-(\Delta\varepsilon^2/4+1)^{1/2}-\Delta\varepsilon/2\leq\kappa\leq min(-\Delta\varepsilon,0)$. For $\Delta\varepsilon= 0$ these transmission bands are reduced to the familiar one of the DNLS, $-1\leq\kappa\leq1$.

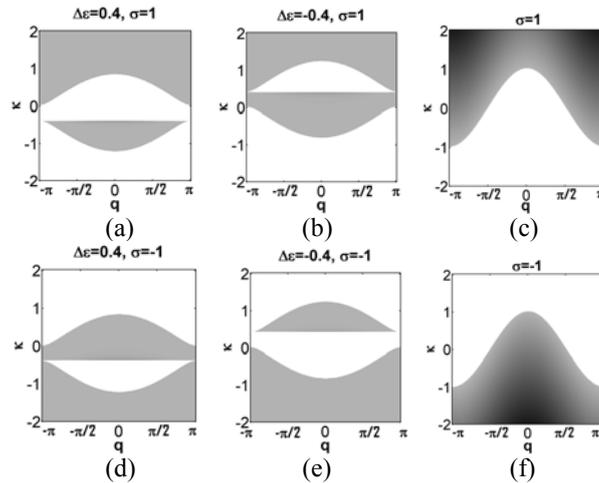

Fig. 2. Diffraction condition for the existence of propagating modes DD(shaded areas) for self-focusing (first row) and self-de focusing (second row) nonlinearity, for interlaced lattices (a, b, d, e) and single nonlinear (c, f).

In the shades areas in Fig. 2, either inside the transmission bands or in the band-gaps, the aforementioned nonlinear diatomic CW solutions can be supported. Furthermore, solitary solutions outside the transmission bands are also expected as we are going to show in Section V. Equation (6) can also be exploited for the investigation of various families of stationary solutions of transverse even-wise periodicity $1$, $2$, $3$ and so on. This is an extensively investigated area in the scientific literature since Eq. (6) is the stationary version of a DNLS for the even sites. However, the propagation dynamics of the interlaced system, Eq.(4), is qualitatively and quantitatively different than a system posed by a single $z$-depended DNLS.

### 3. Modulational instability of the CW solution

Perturbing the amplitude and the phase of a CW solution $v_{2m}=A\exp(-imq)$, $v_{2m+1}=v_{2m}[1+\exp(-iq)]/[2(\kappa+\Delta\varepsilon)]$, with A obeying Eq. (7), as

$$\tilde{\psi}_{2m} = \left[A + u\, e^{i(\Gamma \zeta - 2mQ)} + w^* e^{-i(\Gamma^* \zeta - 2mQ)}\right]e^{i(k_z\zeta - mq)}, \qquad (9)$$

where $u$ and $w$ are functionally independent small amplitude perturbations, $\Gamma$ the propagation constant and $Q$ is their respective spatial frequency. One can easily obtain the following expression for the respective linear response by solving the second linear equation in Eq.(4),

$$\tilde{\psi}_{2m\pm1} = \left[A\frac{1+e^{\mp iq}}{\kappa+\Delta\varepsilon} + u\frac{1+e^{\mp i(q+2Q)}}{\kappa+\Delta\varepsilon+\Gamma}e^{i(\Gamma \zeta - 2mQ)} + w^* \frac{1+e^{\mp i(q-2Q)}}{\kappa+\Delta\varepsilon-\Gamma^*}e^{-i(\Gamma^* \zeta - 2mQ)}\right]\frac{e^{i(k_z\zeta - mq)}}{2} \qquad (10)$$

Utilizing Eqs.(9) and (10) in Eq.(4) and linearizing, one readily obtains,

$$0 = \left\{\left[-\kappa - \Gamma + \frac{1+\cos(q+2Q)}{2(\kappa+\Delta\varepsilon+\Gamma)} + 2\sigma A^2\right]u + \sigma A^2 w\right\}e^{i(\Gamma \zeta - 2mQ)}$$
$$+ \left\{\left[-\kappa + \Gamma^* + \frac{1+\cos(q-2Q)}{2(\kappa+\Delta\varepsilon-\Gamma^*)} + 2\sigma A^2\right]w^* + \sigma A^2 u^*\right\}e^{-i(\Gamma^* \zeta - 2mQ)} \qquad (11)$$

The MI-condition equation can be easily obtained from Eq.(11),

$$\left[-\kappa - \Gamma + \frac{1+\cos(q+2Q)}{2(\kappa+\Delta\varepsilon+\Gamma)} + 2\sigma A^2\right]\left[-\kappa + \Gamma + \frac{1+\cos(q-2Q)}{2(\kappa+\Delta\varepsilon-\Gamma)} + 2\sigma A^2\right] = A^4. \qquad (12)$$

On the basis of Eq. (12), a general remark one can make is that the stability character is unique for fixed values of $\sigma\kappa$ and $\sigma\Delta\varepsilon$: That is, $|Im[\Gamma(\kappa, \Delta\varepsilon)]|$ for focusing media ($\sigma=1$) coincides with $|Im[\Gamma(-\kappa, -\Delta\varepsilon)]|$ for the defocusing ones ($\sigma=-1$). Therefore, without loss of generality, one may fix the signs of $\sigma$ (say, $\sigma=1$) and $\Delta\varepsilon$ and investigate the associated growth rates of the MI over the entire $\kappa$-region $(-\infty<\kappa<\infty)$. In the following, the investigation concerns the upper and the lower edges of the Brillouin zone as well as an indicative case within the that zone for the sake of stressing morphological differences among the edges and the interior.

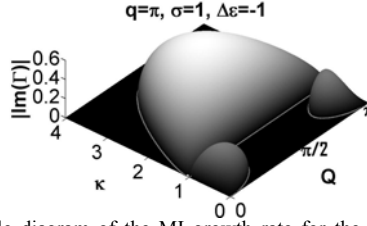

Fig. 3. Gray scale diagram of the MI growth rate for the upper edge of the Brillouin zone ($q=\pi$) as function of the effective propagation constant of the original modulationally unstable CW, $\kappa$, and the spatial frequency of the modulation $Q$ for self-focusing nonlinearity ($\sigma=1$) and $\Delta\varepsilon=-1$.

In Fig. 3, the growth rate of MI, $Im(\Gamma)$, is shown for the upper edge of the Brillouin zone ($q=\pi$) in gray scale diagram as function of the effective propagation constant of the original modulationally unstable CW, $\kappa$, and the spatial frequency of the modulation $Q$. It is worth mentioning that, at this edge, the instability region in $\kappa$ is bounded as shown in Fig. 3. The solid gray curve traces this region which consists of a main lobe extending over the entire range of $Q$ and of two secondary lobes in the vicinity of $Q=0, \pi$. In Fig. 4, the growth rates for two, manifestly unstable choices of $\kappa$, that is, $\kappa=0.5, 2$ (respectively within the secondary and the main lobes) are shown [Fig. 4(a,c)] along with their respective propagation constants, $Re(\Gamma)$ [Fig. 4(b,d)]. In the latter, the unstable modes are shown in gray. The unstable modes within the secondary lobes [Fig. 4(a,b)] are co-moving with the CW, that is, they are characterized $Re(\Gamma)=0$ for MI. In Fig. 5a, the propagation of a perturbed mode corresponding to Fig. 4(c,d) (with $\kappa=2$, that is, within the instability region) is shown. In contrast, in Fig. 5b, the propagation of a perturbed mode with $\kappa=4$ (that is, within the stability region) is also presented. The initial ($\zeta=0$) amplitudes in both cases satisfy the condition Eq.(7). The perturbations on the amplitude and phase $\zeta=0$ are random within $\pm5\%$. The maximum propagation distance ($\zeta=50$) corresponds to about $z_{max} = 2$ cm for an AlGaAs type of configuration as in Fig. 1 ($\lambda=1.55\mu$, TE polarization) with $w_o=10\mu m$, $w_e=2\mu m$, $h_o=5\mu m$, $h_e=1\mu m$, $b=2\mu m$, $L=5\mu m$; this distance is quite large for all practical purposes. For the this edge of the Brillouin zone (upper edge) and according to the aforementioned symmetries of the MI-condition equation [i.e., Eq.(12)] it is clear that for the opposite sign of $\Delta\varepsilon$ (i.e., $\Delta\varepsilon=1$), CW modes exist for $\kappa\geq0$ (as, indeed, it is the case for $\Delta\varepsilon=-1$ and $\kappa\leq0$) and they are all MI-stable for self-focusing media.

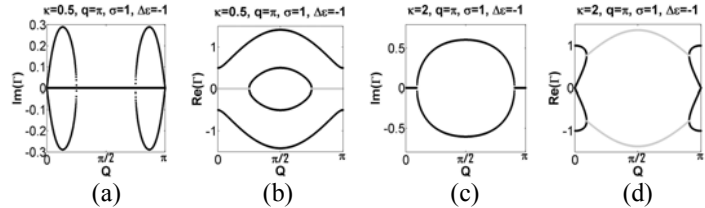

Fig. 4. (a), (c): Growth rates for $\kappa=0.5, 2$ (respectively within the secondary and the main lobes shown in Fig. 3). (b), (d): The respective propagation constants; the unstable modes shown in gray.

For self-defocusing media, on the other hand, and $\Delta\varepsilon=-1$, CW modes exist for $\kappa\leq0$ and they are all MI-stable, while for $\Delta\varepsilon=1$ the behavior is well described by Figs. 3-5 with $\kappa\rightarrow-\kappa$.

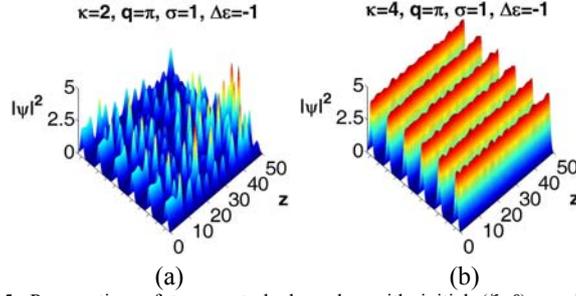

Fig. 5. Propagation of two perturbed modes with initial ($\zeta=0$) amplitudes satisfying Eq.(7). The perturbations on the amplitude and phase at $\zeta=0$ are random within $\pm 5\%$.: (a) corresponding to Fig. 4c-d with $\kappa=2$ (within the instability region); (b) stable propagation of a perturbed mode with $\kappa=4$ (beyond the instability region shown in Fig. 3).

At the lower edge of the Brillouin zone ($q=0$), MI exhibits quite different qualitative features: The instability occurs within a semi-bounded region in $\kappa$. For self-focusing media CW modes exist for $-\Delta\varepsilon/2-[1+(\Delta\varepsilon/2)^2]^{1/2} \leq \kappa < 1$ and $-\Delta\varepsilon/2+[1+(\Delta\varepsilon/2)^2]^{1/2} \leq \kappa < \infty$. The latter semi-bounded region exhibits MI, while the former one is MI-stable. For self-defocusing media, on the other hand, the bounded and the semi-bounded regions become respectively, $1<\kappa\leq-\Delta\varepsilon/2+[1+(\Delta\varepsilon/2)^2]^{1/2}$ which is MI-stable, and $-\infty<\kappa\leq\Delta\varepsilon/2-[1+(\Delta\varepsilon/2)^2]^{1/2}$ where MI is expected to develop. In Fig. 6 the region of instability is shown for self-focusing media and $\Delta\varepsilon=-1$. In Fig. 7, the growth rates for two, manifestly unstable choices of $\kappa$, that is, $\kappa=1.7, 3.5$ (respectively near the edge and in the bulk of the instability region) are shown [Fig. 7(a,c)] along with their respective propagation constants, $Re(\Gamma)$ [Fig. 7(b,d)] where the unstable modes are shown in gray.

It is worth mentioning, at this point, that in the case of a single DNLS lattice the steady state solutions at the lower edge of the Brillouin zone (the so-called unstaggered modes) are always MI-unstable (MI-stable) for self-focusing (self-defocusing) media. However, in the interlaced model in hand, at the lower edge of the Brillouin zone, there always co-exist MI-stable (MI-unstable) CW modes for self-focusing (self-defocusing) media within an isolated region in κ-space. In Figs. 8a-8b, the propagation of two perturbed modes corresponding to Figs. 7a-7d.are shown The perturbations on the amplitude and phase at the launching point ($\zeta=0$) are random within ±5% as previously. In Fig. 8c the propagation of a perturbed mode with $\kappa=0.4$, that is within the stability region, is also shown. The short scale amplitude variations are intrinsic to the particular choice of the amplitude at $\zeta=0$ and they are going to be elucidated in the next Section. The sustained longer scale amplitude variations are due to the random amplitude and phase perturbations initially imposed.

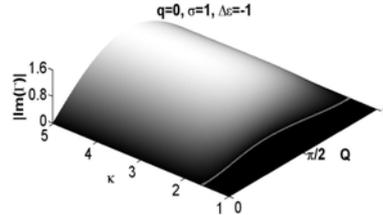

Fig. 6. Gray scale diagram of the MI growth rate for the lower edge of the Brillouin zone ($q=0$) as function of the effective propagation constant of the original modulationally unstable CW, $\kappa$, and the spatial frequency of the modulation $Q$ for self-focusing nonlinearity ($\sigma=1$) and $\Delta\varepsilon=-1$.

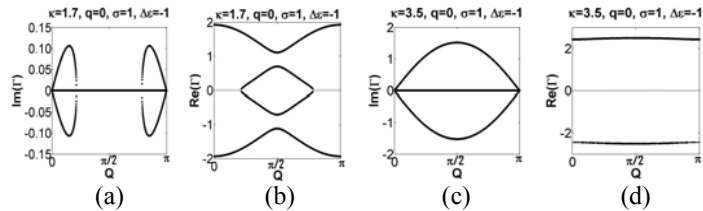

Fig. 7. (a), (c): Growth rates for $\kappa=1.7, 3.5$ (respectively near the edge and in the bulk of the instability region shown in Fig. 6). (b), (d): The respective propagation constants; the unstable modes shown in gray.

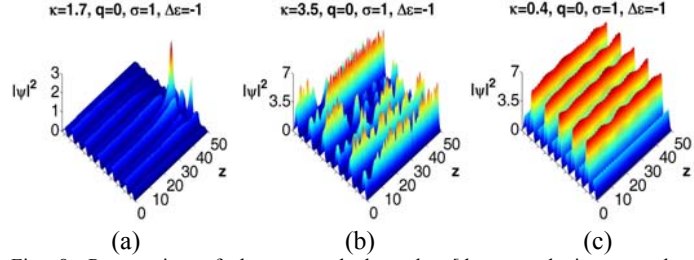

(a) (b) (c)

Fig. 8. Propagation of three perturbed modes [the perturbations on the amplitude and phase at the launching point ($\zeta=0$) are random within ±5%.]: (a), (b): unstable, corresponding to Fig. 7; (c) stable propagation of a perturbed mode with $\kappa=0.4$ (that is, beyond the instability region shown in Fig. 6).

Within the Brillouin zone MI attains different characteristics. To qualitatively illustrate this point two examples are provided: For self-focusing media, $\Delta\varepsilon=-1$ and $q=2\pi/3$ (that is near the upper edge), the instability remains bounded (as in the case $q=\pi$) and it is shown in Fig. 9. The solid gray curve traces this region which consists of a main lobe extending over the entire range of $Q$ while the two secondary lobes in the vicinity of $Q=0, \pi$, which previously characterized the upper edge of the Brillouin zone (in Fig. 3), now coalesce in a single disjoint region of instability. In Fig. 10, the growth rates for three manifestly unstable choices of $\kappa$, that is, $\kappa=0, 1.5, 2.0$ (one within the secondary region and two within the main one) are

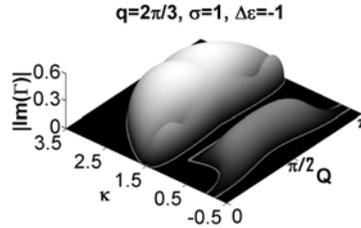

Fig. 9. Gray scale diagram of the MI growth rate within the Brillouin zone [near the upper edge of ($q=2\pi/3$)] as function of the effective propagation constant of the original modulationally unstable CW, $\kappa$, and the spatial frequency of the modulation $Q$ for self-focusing nonlinearity ($\sigma=1$) and $\Delta\varepsilon=-1$.

shown (first row) along with their respective propagation constants, $Re(\Gamma)$ (second row). In the latter, the unstable modes are shown in gray. In Figs. 11a-11b, the propagation of two perturbed modes corresponding to the first two columns of Fig. 10 ($\kappa=0, 1.5$) are shown. In contrast, in Fig. 11c the propagation of a perturbed mode with $\kappa=5$ (that is, well inside the

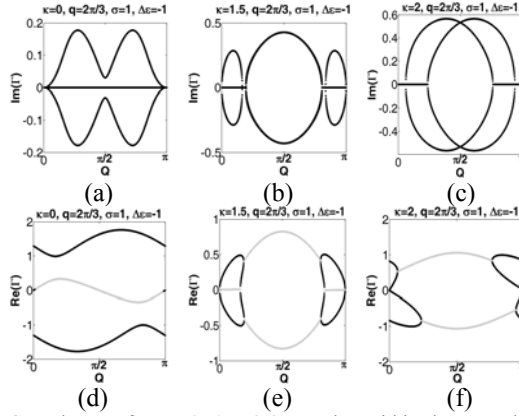

Fig. 10. Growth rates for, $\kappa=0, 1.5, 2$ (one value within the secondary region shown in Fig. 9 and two within the main one) are shown (first row) along with their respective propagation constants (second row). In the latter, the unstable modes are shown in gray.

stability region) is shown. Again, in all these cases, the initial ($\zeta=0$) amplitudes satisfy the condition Eq. (7) while the random perturbations on their amplitude and phase $\zeta=0$ are within $\pm 5\%$. On the other hand, for $\Delta\varepsilon=1$ and for self-focusing media MI-stable CW modes now exist for all positive $\kappa$'s in the semi-bounded region $[1/4+(\Delta\varepsilon/2)^2]^{1/2}-\Delta\varepsilon/2\leq\kappa$ (the lower bound was $0$ for $q=\pi$). However, a bounded second region of existence appears for $-\Delta\varepsilon/2-[1/4+(\Delta\varepsilon/2)^2]^{1/2}\leq\kappa<-1$. All CW modes in the latter region are weakly MI-unstable [that is, $|Im(\Gamma)|$ is comparatively very small in this region]. For self-defocusing media, stable CW modes exist for all negative values of $\kappa$ in the semi-bounded region $\kappa\leq-[1/4+(\Delta\varepsilon/2)^2]^{1/2}-\Delta\varepsilon/2$ while unstable ones appear in the positive bounded region $1<\kappa\leq-\Delta\varepsilon/2+[1/4+(\Delta\varepsilon/2)^2]^{1/2}$. In the second example (again, $q=2\pi/3$), the instability region for a self-defocusing medium and

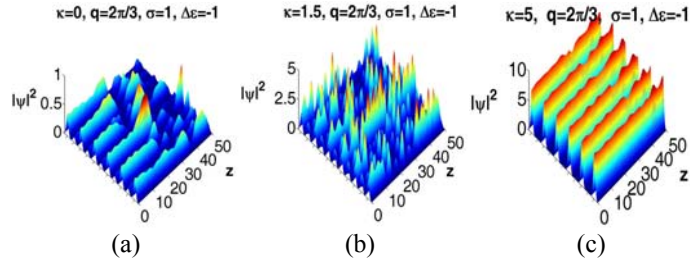

Fig. 11. Propagation of three perturbed modes [the perturbations on the amplitude and phase at the launching point ($\zeta=0$) are random within $\pm 5\%$.]: (a), (b): unstable, corresponding to the first and second column of Fig. 10; (c) stable propagation of perturbed mode with $\kappa=5$ (that is, beyond the instability region shown in Fig. 9).

$\Delta\varepsilon=-1$ is shown (Fig. 12). This case is equivalent (upon $\kappa \longrightarrow -\kappa$) to a case of a self-focusing medium, as in the previous example, with, $\Delta\varepsilon=1$. In Fig. 13, the growth rates an unstable choice of $\kappa$, ($\kappa=1.12$) is shown along with the respective propagation constants, $Re(\Gamma)$. Additionally, in Fig.14, the propagation of MI-stable mode (for a negative value of $\kappa$ in this particular case) is contrasted with an unstable one (right sub-figure). Since the instability is weak, the propagation distance is set to a very large value in order to allow the instability to develop. The random perturbations on the amplitude and phase at $\zeta=0$ are also set within the quite wide margin of $\pm 25\%$. The short scale amplitude variations of the stable mode are both due to this quite strong initial perturbations and to the particular choice of the amplitude at $\zeta=0$ as well.

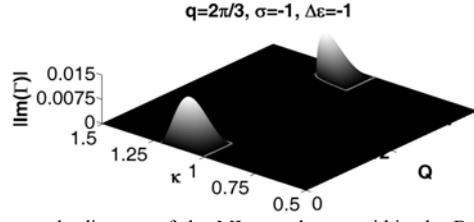

Fig. 12. Gray scale diagram of the MI growth rate within the Brillouin zone ($q=2\pi/3$) as function of the effective propagation constant of the original modulationally unstable CW, $\kappa$, and the spatial frequency of the modulation $Q$ for self-defocusing nonlinearity ($\sigma=-1$) and $\Delta\varepsilon=-1$.

## 4. Propagation dynamics of the CW solutions at the lower edge of the Brillouin zone

Starting from the original system, Eq. (4), for a continuous wave solutions (CW) of zero transverse momentum [that is, CW modes launched transversely to the $(x,y)$ plane, or, equivalently, those with $q=0$] one may set,

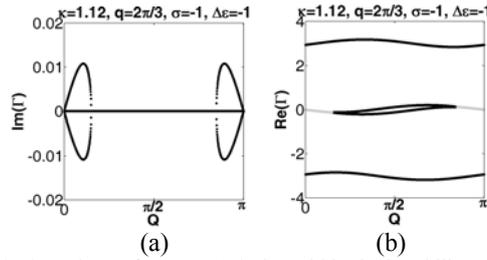

(a)  (b)

Fig. 13. (a): Growth rate for, $\kappa=1.12$ (value within the instability region shown in Fig. 12); (b): its respective propagation constant. In the latter, the unstable modes are shown in gray.

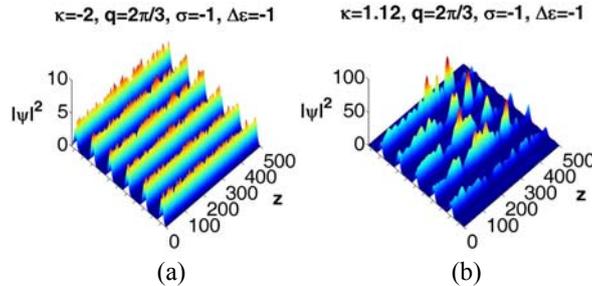

(a)  (b)

Fig. 14. Propagation of two perturbed modes with initial ($\zeta=0$) amplitudes satisfying Eq.(7). The perturbations on the amplitude and phase at $\zeta=0$ are random within $\pm 5\%$.: (a) stable propagation of a perturbed mode with $\kappa=-2$ (beyond the instability region shown in Fig. 12); (b). unstable propagation of a perturbed mode with $\kappa=1.12$ corresponding to Fig. 13. Notice that the propagation distance is quite large.

$$\psi_{2m+1} = \psi_{odd}(\zeta) = \rho_{odd}(\zeta)\exp\left[i\phi_{odd}(\zeta)\right], \quad \psi_{2m} = \psi_{even}(\zeta) = \rho_{even}(\zeta)\exp\left[i\phi_{even}(\zeta)\right]. \quad (13)$$

Thus, the original system, in polar form, becomes,

$$\frac{d\rho_{even}}{d\zeta} = \rho_{odd} \sin \Delta\phi, \qquad \frac{d\rho_{odd}}{d\zeta} = -\rho_{even} \sin \Delta\phi$$

$$\frac{d\Delta\phi}{d\zeta} = \left(\frac{\rho_{odd}}{\rho_{even}} - \frac{\rho_{even}}{\rho_{odd}}\right)\cos\Delta\phi + \Delta\varepsilon + \sigma\rho_{even}^2 \qquad (14)$$

$$\Delta\phi \equiv \phi_{even} - \phi_{odd}, \quad \Delta\varepsilon \equiv \varepsilon_{even} - \varepsilon_{odd}, \quad S \equiv \rho_{even}^2 + \rho_{odd}^2 = \text{constant} \equiv S_0 > 0$$

It is clear that the combined intensity is a constant of the motion ($S_0$). By introducing a new combined dynamical variable $D = \rho^2_{even} - \rho^2_{odd}$, the previous system becomes,

$$\frac{d\Delta\phi}{d\zeta} = -\frac{D}{\rho_{even}\rho_{odd}}\cos\Delta\phi + \Delta\varepsilon + \sigma\rho_{even}^2$$

$$\frac{dD}{d\zeta} = 4\rho_{even}\rho_{odd}\sin\Delta\phi \qquad (15)$$

$$\rho_{even}^2 = \frac{S+D}{2}, \qquad \rho_{odd}^2 = \frac{S-D}{2}$$

By eliminating $\rho_{odd}$ and $\rho_{even}$ one finally gets,

$$\frac{d\Delta\phi}{d\zeta} = \frac{-2D}{\sqrt{S_0^2 - D^2}}\cos\Delta\phi + \Delta\varepsilon + \frac{\sigma}{2}(S_0 + D), \quad \frac{dD}{d\zeta} = 2\sqrt{S_0^2 - D^2}\sin\Delta\phi \qquad (16)$$

Notice that, for the case of a single lattice, this system reduces to the following one:

$$\frac{d\phi}{d\zeta} = \varepsilon + \sigma\rho^2, \quad \rho^2 = \frac{S_0}{2} = \text{constant}, \qquad \psi(\zeta) = \rho\exp\left[i(\varepsilon + \sigma\rho^2)\zeta\right] \qquad (17)$$

The fixed points [$D = D(\zeta=0)$, $\Delta\varphi_0 = \Delta\varphi(\zeta=0) = 0, \pi$] of the system in Eq. (16) can be found by solving the equation:

$$\Delta\varepsilon = \frac{2D}{\sqrt{S_0^2 - D}}\cos\Delta\phi_0 - \frac{\sigma}{2}(S_0 + D) \qquad (18)$$

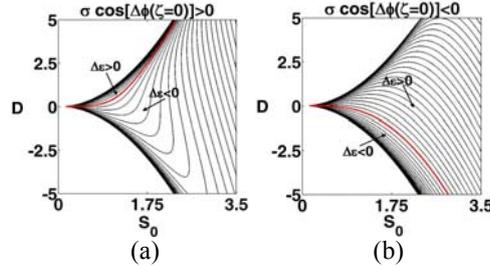

Fig. 15. $D$-$S_0$ diagram of the fixed points loci for various values of $\Delta\varepsilon$. The red line corresponds to $\Delta\varepsilon = 0$. (a): for $\sigma\cos[\Delta\varphi(\zeta=0)] > 0$ [that is, for focusing (defocusing) nonlinearity and phase difference $0$ ($\pi$)] there exist three fixed points beyond a particular value of the constant of propagation, $S_0$; (b): for $\sigma\cos[\Delta\varphi(\zeta=0)] < 0$ [that is, for focusing (defocusing) nonlinearity and phase difference $\pi$ ($0$)] $D(S_0)$ there is always a single fixed point.

For fixed values of $\Delta\varepsilon$ and $S_0$ (constant of propagation) one or three fixed points exist (i.e., one or three distinct values of $D$) as shown in Fig. 15. The transition from one to three occurs only for $\sigma\cos[\Delta\varphi(\zeta=0)]>0$ [that is, for focusing (defocusing) nonlinearity and phase difference $0$ ($\pi$), as in Fig. 15a where beyond some value of $S_0$, $D(S_0)$ becomes multi-valued function]. In Fig. 15 contours of constant $\Delta\varepsilon$ are shown. The red line corresponds to $\Delta\varepsilon=0$. One may easily show that the value of $D$ at the transition point can then be expressed in terms of $S_0$ as follows:

$$\frac{\partial \Delta\varepsilon}{\partial D} \bigg/ \frac{\partial \Delta\varepsilon}{\partial S_0} = 0 \Rightarrow D = \pm S_0^{2/3}\sqrt{S_0^{2/3} - 2^{4/3}}$$

The dynamical system exhibits Hamiltonian structure:

$$\mathcal{H}(D,\Delta\phi) = 2\sqrt{S_0^2 - D^2}\cos\Delta\phi + \left(\Delta\varepsilon + \frac{\sigma}{2}S_0\right)D + \frac{\sigma}{4}D^2 = \text{constant} = H_0$$
$$\frac{d\Delta\phi}{d\zeta} = \frac{\partial \mathcal{H}}{\partial D}, \quad \frac{dD}{d\zeta} = -\frac{\partial \mathcal{H}}{\partial \Delta\phi} \tag{19}$$

Therefore, the system can be integrated in terms of elliptic functions,

$$\frac{\zeta}{4} = \pm\int_{D(\zeta=0)}^{D(\zeta)} \frac{dw}{\sqrt{a_0 + a_1 w - a_2 w^2 - a_3 w^3 - w^4}}$$
$$a_0 = 16(4S_0^2 - H_0^2), \quad a_1 = 16H_0\sigma(S_0 + 2\sigma\Delta\varepsilon) \tag{20}$$
$$a_2 = 4S_0(S_0 + 4\sigma\Delta\varepsilon) + 8(8 + 2\Delta\varepsilon^2 - \sigma H_0), \quad a_3 = 4(S_0 + 2\sigma\Delta\varepsilon)$$

where the sign in front of the integral is determined by $\Delta\varphi(\zeta=0)$. Only periodic or bounded Jacobi elliptic functions are acceptable. Therefore, proper conditions on the choice of $\Delta\varphi(\zeta=0)$, $\rho_{even}(\zeta=0)$, $\rho_{odd}(\zeta=0)$ must be met. However there is plenty of flexibility available because one has three parameters to adjust.

The stability of the fixed points involved for a particular choice of the parameters $\Delta\varepsilon$ and $S_0$ is controlled via the sign of the parameter $\Delta$ ($\Delta>0$: stable, $\Delta<0$: unstable) at the fixed points,

$$\Delta = \frac{\partial^2 \mathcal{H}}{\partial D^2}\frac{\partial^2 \mathcal{H}}{\partial \Delta\phi^2} - \left(\frac{\partial^2 \mathcal{H}}{\partial D \partial \Delta\phi}\right)^2 \tag{21}$$

In the Figs. 16a-16f, the parameter regions the nature and the number of fixed points are depicted. For zero phase difference between the odd and the even sites ($\Delta\varphi=0$, first column) and for focusing nonlinearity there exist both stable (elliptic) and unstable (hyperbolic) fixed points while for defocusing nonlinearity there exist only one elliptic fixed point. On the other hand, for phase difference $\Delta\varphi=\pi$ (second column) for focusing nonlinearity there exist only one elliptic point, while for defocusing nonlinearity there exist both stable (elliptic) and unstable (hyperbolic) fixed points. In the vicinity of the elliptic points stable and periodic in $\zeta$ CW solutions exist possessing not necessarily the same amplitude in the odd and even sites. This periodic in $\zeta$ behavior is visible, for example, in Fig. 8c: Due to the random perturbation imposed at $\zeta=0$, the CW modes exhibit short scale stable variations in $\zeta$ since the system has been slightly displaced from its stable fixed point configuration.

In Fig. 17 a case of three fixed points for $\sigma=1$, $\Delta\varphi(\zeta=0)=0$, $\Delta\varepsilon=-2.7$, two elliptic (O, red and green) and one hyperbolic (X) is shown. The constant of propagation is set to $S_0=5.76$, while the other constant of propagation, namely the Hamiltonian, $H_0$, acquires the values indicated. For $11.431<H_0<11.648$ two families of $\zeta$-periodic modes, separately encircling the two elliptic points, exist. For $11.648<H_0<12.99$, only one family of $\zeta$-periodic modes exists encircling the upper elliptic point. On the other hand, for $H_0<11.431$ there is a new family of $\zeta$-periodic modes which encircles all three fixed points; the red contour is the separatrix which separates the latter family for the innermost $\zeta$-periodic modes. It should be emphasized that the separatrix corresponds to a family of unstable modes with infinite period in $\zeta$. If they were stable, they could preserve indefinitely the initially chosen $\Delta\varphi$ and $D$ or, equivalently, the same amplitudes in even and odd lattice sides and the same difference in their respective phases. One final important remark: There are no amplitude-only-stationary CW solutions with $q=0$, that is, solutions of the form as in Eq. (13) with constant amplitude and periodic phase variation. That is, the single lattice respective modes, as in Eq. (17) cannot be extended to the interlaced lattice model.

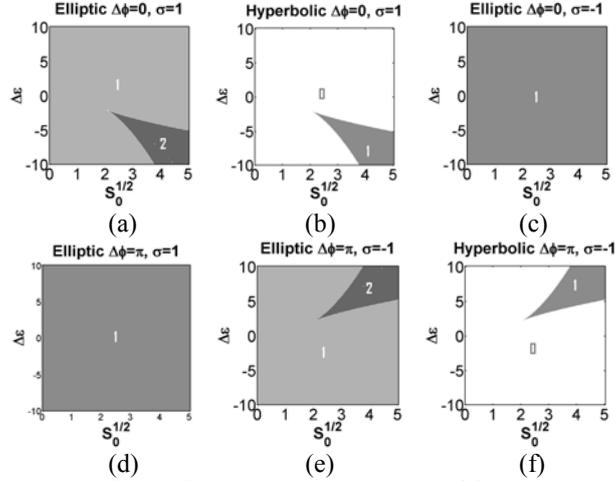

Fig. 16. $\Delta\varepsilon$-$S_0$ diagram for the nature and the number of fixed points. (a), (b): $\Delta\varphi=0$, $\sigma=1$ (both elliptic and hyperbolic fixed points exist); (c) $\Delta\varphi=0$, $\sigma=-1$ (only elliptic fixed points exist); (d) $\Delta\varphi=\pi$, $\sigma=1$ (only elliptic fixed points exist); (e), (f): $\Delta\varphi=\pi$, $\sigma=-1$ (only hyperbolic point exist).

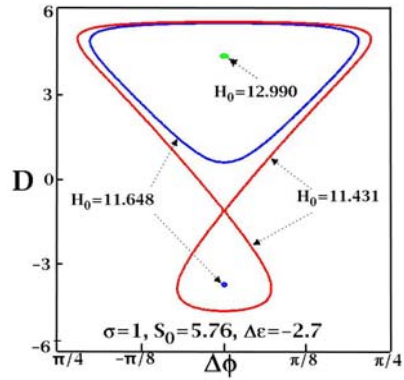

Fig. 17. A case of three fixed points for $\sigma=1$, $\Delta\varphi(\zeta=0)=0$, $\Delta\varepsilon=-2.7$, two elliptic (O, red and green) and one hyperbolic (X). The constant of propagation is set to $S_0=5.76$, while the other constant of propagation, namely the Hamiltonian, $H_0$, acquires the values indicated. The red contour is the separatrix.

For CW solutions with nonzero transverse momentum ($q\neq 0$) a multitude of possibilities opens up among which there exist modes of transverse period 1, 2, etc in each sub-lattice. However, as far as the particular underlying dynamical system is concerned several degrees of freedom may now be involved that render the system non-integrable in the general case. This, of course, goes beyond the scope of the present work.

## 5. Solitons

We have shown in Section II that the band structure of the interlaced lattice system is different than the one for the single nonlinear lattice case: There is a third bounded band-gap between the two semi-infinite band-gaps, that is, located between the acoustic and the optical branches. Because of this extra gap, the comparison with the DNLS of the single lattice becomes rather tedious and, therefore, in this Section we will limit ourselves in a few basic remarks. Characteristic cases of solitary modes with propagation constants inside these three band gaps will be presented. All solitary solutions have been found via the Newton's iteration method employed on Eq.(6). Furthermore, in all cases shown, the maximum propagation distance ($\zeta=100$) corresponds to about $z_{max} = 4$ cm [for an AlGaAs type of configuration as in Fig. 1 ($\lambda=1.55\mu$, TE polarization) with $w_o=10\mu m$, $w_e=2\mu m$, $h_o=5\mu m$, $h_e=1\mu m$, $b=2\mu m$, $L=5\mu m$], quite large for all practical purposes.

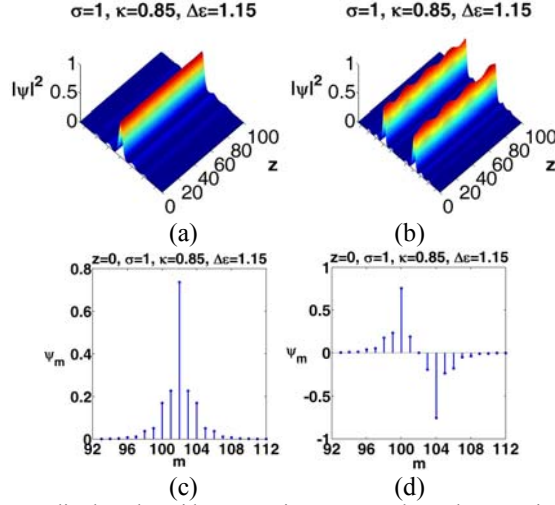

Fig. 18. Localized modes with propagation constant above the acoustic branch ($\kappa=0.85$) for $\Delta\varepsilon=1.15$ and self-focusing nonlinearity. (a,b): intensity profile upon propagation, (c,d): initial (at $\zeta=0$) profiles; The latter are perturbed randomly in amplitude and phase within 10%.

In Fig. 18 two cases with propagation constant located in the upper semi-infinite band gap (that is, above the acoustic branch) are shown: The case in Fig. 18(a,c) corresponds to an in-phase fundamental solution, while the one in Fig. 18(b,d) represents a bound state with dipole-like initial profile. Both cases refer to a self-focusing nonlinearity with $\kappa=0.8$ and $\Delta\varepsilon=1.15$. The transmission bands for this choice of $\Delta\varepsilon$, are $-1.73<\kappa<-1.15$ and $0<\kappa<0.58$). In both cases, the initial (at $\zeta=0$) profile is perturbed randomly in amplitude and phase within 10%.

In the lower semi-infinite band gap (that is, below the optical branch) one expects localized solutions only for self-defocusing nonlinearity ($\sigma=-1$). Indeed, in Fig. 19 two such cases with propagation constant, $\kappa=-3$, located in the lower semi-infinite band gap (for the same choice of $\Delta\varepsilon$ as in Fig. 18) are shown. The case in Fig. 19(a,c) corresponds to an in-phase fundamental solution, while the one in Fig. 19(b,d) represents a bound state with dipole-like staggered initial profile. In both cases, the initial (at $\zeta=0$) profile is perturbed randomly in amplitude and phase within 5%.

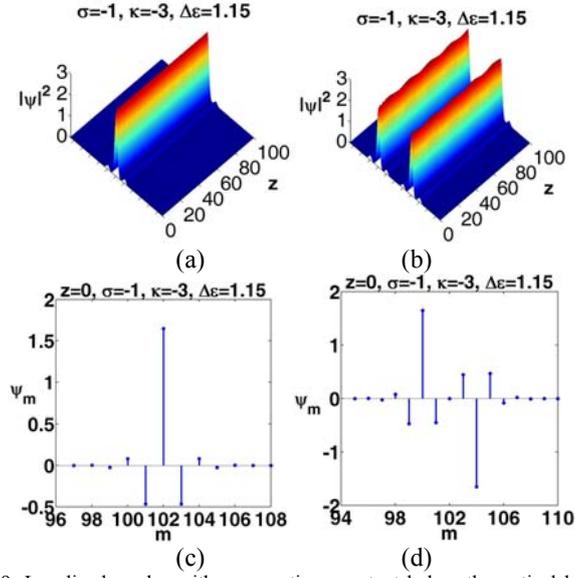

Fig. 19. Localized modes with propagation constant below the optical branch ($\kappa=-3$) for $\Delta\varepsilon=1.15$ and self-defocusing nonlinearity. (a,b): intensity profile upon propagation, (c,d): initial (at $\zeta=0$) profiles; The latter are perturbed randomly in amplitude and phase within 5%.

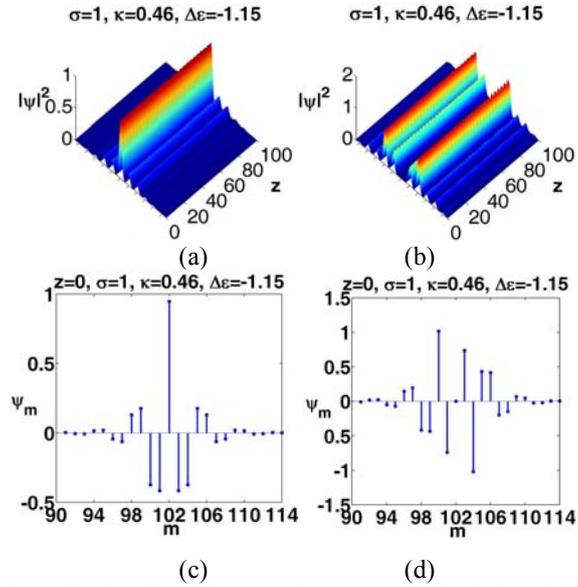

Fig. 20. Localized modes with propagation constant inside the bounded band-gap ($\kappa=0.46$) for $\Delta\varepsilon=-1.15$ and self-focusing nonlinearity. (a,b): intensity profile upon propagation, (c,d): initial (at $\zeta=0$) profiles; The latter are perturbed randomly in amplitude and phase within 5%.

Finally, two characteristic cases with $\kappa$ laying inside the extra gap are shown in Fig. 20 for self-focusing nonlinearity ($\sigma=1$) and $\Delta\varepsilon=-1.15$ (for this choice of $\Delta\varepsilon$ the transmission bands are $-0.58<\kappa<0$ and $1.15<\kappa<1.73$). The propagation constant is $\kappa=0.46$. The case in Fig. 20(a,c) corresponds to a staggered type of solution (apart of the central sites, the sign flips for even and odd ones separately), while the one in Fig. 20(b,d) represents a bound state with an initial profile which is dipole-like and staggered too. In both cases, the initial (at $\zeta=0$) profile is perturbed randomly in amplitude and phase within 5%. Comparatively speaking (Fig. 20 vs. Fig. 19, though for different sign in the nonlinearity) the numerical investigation have shown that the extra gap is populated by more structurally rich localized modes, that is, modes with several sites appreciably active

## 6. Summary and conclusions

In this Section the main observations and new features that the interlaced lattice model introduces are summarized. The model is basically a diatomic one and it is characterized by two coupled DNLSEs. However, due to the effectively linear character of the second DNLSE, it was possible to analytically investigate the behavior of the system to a considerable extent. However, the present investigation was not exhaustive since the problem in hand has a multitude of aspects which worth further investigation. However, the findings and the conclusions of this study provide a firm analytical basis for further investigation.

An important design parameter for the lattice which can be suitably chosen is the normalized relative detuning $\Delta\varepsilon$ which dictates the existence of two distinct CW modes at each constituent lattice. For self-focusing media CW modes exist for values of the effective propagation constant, $\kappa$, of the CW inside the two disjoint regions $-\Delta\varepsilon/2-[1+(\Delta\varepsilon/2)^2]^{1/2} \leq \kappa < 1$ and $-\Delta\varepsilon/2+[1+(\Delta\varepsilon/2)^2]^{1/2} \leq \kappa < \infty$. For self-defocusing media, on the other hand, these regions are $1 < \kappa \leq -\Delta\varepsilon/2+[1+(\Delta\varepsilon/2)^2]^{1/2}$ and $-\infty < \kappa \leq -\Delta\varepsilon/2-[1+(\Delta\varepsilon/2)^2]^{1/2}$.

As far as the modulational instability of the CW modes is concerned, several observations can be made: The stability character is unique for fixed values of the $\sigma\kappa$ and $\sigma\Delta\varepsilon$ ($\sigma$ is the sign of the cubic nonlinearity). Therefore, without loss of generality, one may fix the signs of the cubic nonlinearity and the sign of the design parameter $\Delta\varepsilon$ and investigate the associated growth rates of the MI over the entire $\kappa$-region $(-\infty<\kappa<\infty)$. At the upper edge of the Brillouin zone, namely for Bloch momentum $q=\pi$, the instability region in $\kappa$ is bounded. This region consists of a main lobe extending over the entire range of the values of modulation frequency, Q, and of two secondary lobes in the vicinity of $Q=0, \pi$. The unstable modes within the secondary lobes are co-moving with the CW, that is, their respective real part of the exponent $\Gamma$ for MI is zero. On the other hand, at the lower edge of the Brillouin zone $(q=0)$, MI exhibits quite different qualitative features: For self-focusing media, the instability occurs only within the semi-bounded region in κ. while the bounded one is MI-stable. The same applies for self-defocusing media where MI is expected to develop within the respective semi-bounded region. It is worth mentioning that, in the case of a single DNLS lattice, the steady state solutions at the lower edge of the Brillouin zone (the so-called unstaggered modes) are always MI-unstable (MI-stable) for self-focusing (self-defocusing) media. In contrast, in the interlaced model in hand, at the lower edge of the Brillouin zone, there always co-exist MI-stable (MI-unstable) CW modes for self-focusing (self-defocusing) media within an isolated region in κ-space. Within the Brillouin zone MI is modified accordingly: For instance, for self-focusing media and near the upper edge, the instability remains bounded in the main lobe, while the two aforementioned secondary lobes coalesce in a single disjoint region of instability. However, near the same edge, for self defocusing media, MI spreads now inside the bounded region as well.

The propagation dynamics of CW modes at the lower edge of the Brillouin zone has also been investigated since, at this edge, the behavior is amenable to a rather straightforward analytical treatment. Additionally, from the practical point of view, these modes are fundamental: That is, they represent CW modes launched transversely onto the face of the lattice sides. It is worth mentioning that there are no amplitude-only-stationary CW solutions in this case, that is, solutions with constant amplitude and periodic phase variation. The corresponding dynamical system has two evolving in $z$ dynamical variables, namely, the common phase difference among the sites of the constituent lattices and the difference of their respective squared amplitudes (i.e., their intensities). The system exhibits Hamiltonian structure, it possesses two constants of the motion (propagation invariants), namely the

Hamiltonian function itself [Eq.(19)] and the sum of the squares of the respective amplitudes. The equations of motion are readily integrated and, therefore the dynamical variables are expressed in terms of elliptic functions. The stability of these CW modes is tightly associated with the existence and the character of the fixed point of this underlying dynamical system. The character and the number of these points is controlled by the design parameter $\Delta\varepsilon$, which dictates the MI as we have already shown, and the common sum of the intensities of adjacent sites (as illustrated in Fig. 16). It is worth mentioning that, for focusing (defocusing) nonlinearity and phase difference $0$ ($\pi$) [that is, for in-phase (staggered-type)] among adjacent sites, there exist three fixed points beyond a particular value of this common sum (which is conserved upon propagation). In such a case, two of the fixed points are elliptic and one is hyperbolic. In the vicinity of the elliptic points there exist stable and periodic in $z$ CW modes possessing not necessarily the same amplitude in the odd and even sites. On the other hand, through the hyperbolic point passes a separatrix. The latter separates three families of periodic in $z$ modes: Periodic CWs which encircle all three fixed points and CWs in the vicinity of each elliptic point. The former one represents modes with respective amplitudes and phases in adjacent sites varying in a considerably wider range.

The existence and the form of stable localized solutions were also investigated. However, the presence of an extra gap (controlled by the design parameter $\Delta\varepsilon$), in a region which is basically a transmission region for the single DNLS lattice, renders the comparison with the latter quite tedious. Therefore, the investigation was limited to a few characteristic cases in all three band-gaps. In the upper semi-infinite band-gap, that is, beyond the acoustic branch, in-phase discrete solitons were found for self-focusing materials. In this band-gap, there exist a variety of both fundamental (center-peaked) localized solutions as well as dipole ones. Inside the extra gap, there also exist both previous types. They are, however of a staggered sort of type. None in phase stable localized modes were found in this gap. Finally, in the semi-infinite band-gap below the optical branch, localized modes were found for self-defocusing media, as expected. They are, too, of staggered sort of type, fundamental or dipole ones. In conclusion, by choosing the propagation constant $\kappa$, and accordingly adjusting the design parameter $\Delta\varepsilon$, a multitude of localized modes can, thus, be generated. There is evidence that the extra gap is populated by more rich structurally (several sites are active) localized modes.

The appearance and the controllability, by design, of the extra gap in the interlaced lattice model presented in this work, lead to a richness in both the form and the dynamics of the stable modes which are supported. These attributes, the interlaced lattice system possesses, may decisively contribute to an enhanced functionality of devices of the type discussed.

.